\documentclass[pra,amsmath,amsfonts,amssymb,showpacs]{revtex4} %,nofootinbib
\usepackage{graphicx}
\def\beq{\begin{equation}}
\def\eeq{\end{equation}}
\def\bea{\begin{eqnarray}}
\def\eea{\end{eqnarray}}

\newcommand{\be}{\begin{equation}}
\newcommand{\ee}{\end{equation}}

\newcommand{\ed}{\end{document}}

\newcommand{\bi}{\begin{itemize}}
\newcommand{\ei}{\end{itemize}}

\newcommand{\bce}{\begin{center}}
\newcommand{\ece}{\end{center}}
\begin{document}

 \vspace*{3cm}

\title{Superdense Coding with Uniformly Accelerated Particle}

\author{Mehrnoosh Farahmand}
\affiliation{Department of Physics, University of Mohaghegh
Ardabili,  P.O. Box 179, Ardabil, Iran}
%\email{mohammadzadeh@uma.ac.ir}

\author{Hosein Mohammadzadeh}
\affiliation{Department of Physics, University of Mohaghegh
Ardabili,  P.O. Box 179, Ardabil, Iran}
%\email{mohammadzadeh@uma.ac.ir}

\author{Hossein Mehri-Dehnavi}
\affiliation{Department of Physics, Babol University of Technology,
Babol, 47148-71167, Iran}
%\email{mehri@nit.ac.ir}

\author{Robabeh Rahimi}
\affiliation{Department of Physics, Science and Research Branch of
Islamic Azad University, Tehran, Iran}
%\email{mohammadzadeh@uma.ac.ir}


\begin{abstract}
We study superdense coding with uniformly accelerated particle in
single mode approximation and beyond single mode approximation. We
use four different functions, the capacity of superdense coding,
negativity, discord and the probability of success for evaluating
the final results. In single mode approximation, all the four
functions behave as expected, however in beyond  single mode
approximation, except the probability of success, the other three
functions represent peculiar behaviors at least for special ranges
where the beyond single mode approximation is strong.
\end{abstract}
 \maketitle
%\keywords{Superdense coding, Relativistic quantum information,
%Beyond single mode approximation, Entanglement, Discord, Capacity of
%superdense coding}
%%%%%%%%%%%%%%%%%%%%%%%%%%%%%%%%%%%%%%%%%%%%%%%%%%%%%
\section{Introduction}
%%%%%%%%%%%%%%%%%%%%%%%%%%%%%%%%%%%%%%%%%%%%%%%%%%%%%
 Two particles, even being far away from each other, can be correlated
as a result of existing nonclassical correlation and entanglement in
between them. Theoretical studies and experimental investigations of
entanglement and nonclassical correlation have been main topics for
groups of researchers  \cite{Monroe,Bennett,Ursin1,Mehri1,Mehri2}.
In the process of so called superdense coding \cite{Bennentt} two
classical bits of information are transferred by sending only one
quantum bit, qubit. The original superdense coding process begins
with a pair of entangled two-level particles being shared between
Alice, sender, and Bob, receiver. An EPR pair \cite{EPR} is used as
a maximally entangled state. We have four orthonormal EPR states
which can be written as
    \bea\label{in}
        |\varphi_{\alpha\beta}\rangle_{AB}=\frac{1}{\sqrt{2}}\left\{|0\rangle_{A}|\alpha\rangle_{B}+(-1)^{\beta}|1\rangle_{A}|\overline{\alpha}\rangle_{B}\right\},
    \eea
where $\alpha,\beta=\{0,1\}$, $\overline{\alpha}=1-\alpha$ and
subscripts $A$ and $B$ denote Alice's qubit and Bob's qubit,
respectively.

 Let us assume, without loss of generality, that Alice and Bob share the state
$|\varphi_{00}\rangle_{AB}, ~ \alpha=\beta=0$. Alice has a two-bit
message that she wants to send it to Bob. The classical two-bit
message can be one of the forms $ij=\{00, 01, 10, 11\}$. Alice first
operates one of the four unitary operators $U_{ij}=Z^{j}X^{i}$ on
her qubit. $X$ and $Z$ are Pauli operators. Consequently, the
initial EPR pair changes to one of the four orthonormal EPR states,
$|\varphi_{ij}\rangle$, i.e. the original EPR state is encoded by
the message, $ij$. Then, Alice sends her manipulated qubit to Bob,
who performs a measurement in the Bell-basis, that yields the
classical message, $ij$. Superdense coding has been experimentally
implemented \cite{Mattle,Fang,Rahimi,Jietai,Mizuno}.

\begin{figure}
  \includegraphics[width=0.5\columnwidth]{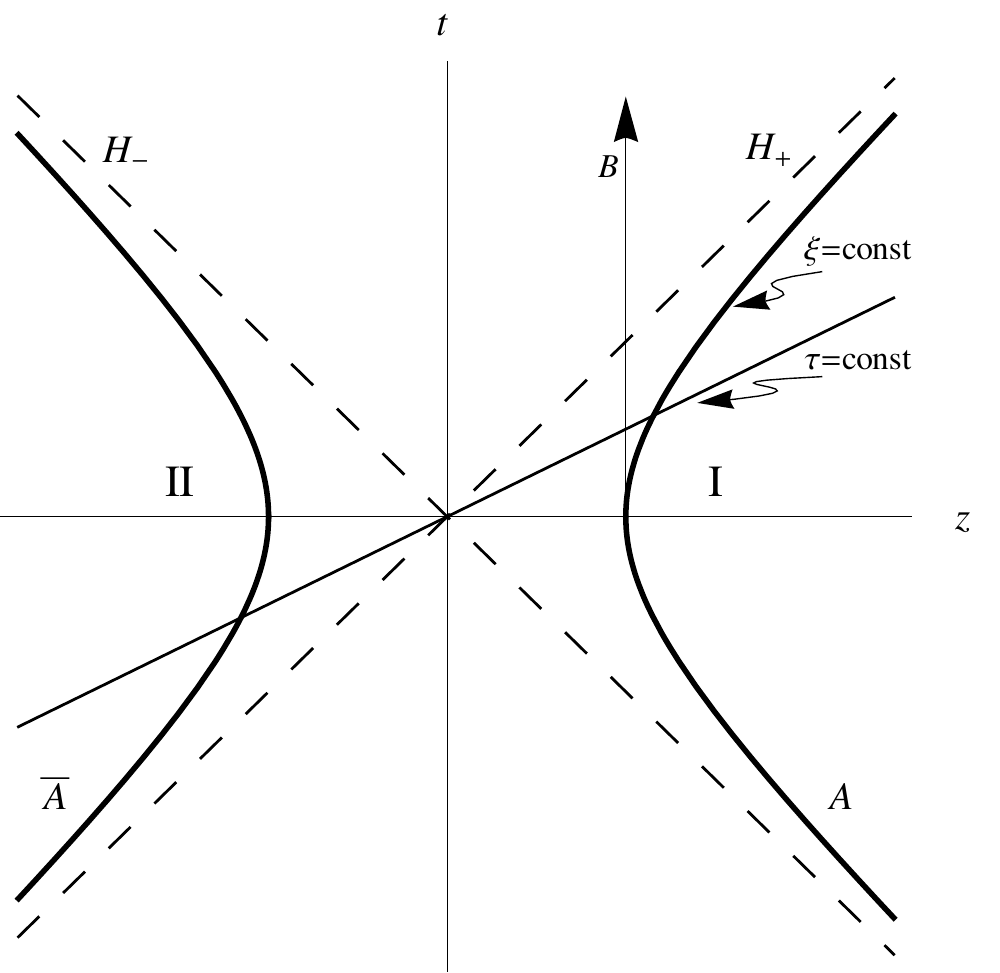}\\
  \caption{Minkowski diagram for Alice and Bob. Bob is stationary and Alice travels with constant acceleration and is moving along the hyperbola in region I while fictitious observer anti-Alice moving along a corresponding hyperbola in region II. Bob will cross from the horizons $H_{\pm}$ at his finite Minkowski time $t_{A}$. After this time Alice's signals can just across from $H_{+}$ and arrives to Bob. }\label{Fig0}
\end{figure}

 In this paper, we suppose two particles denoted as Alice and Bob.
Alice is accelerated while Bob stays inertial. Therefore, one can
say that Alice has constant acceleration with respect to Bob in the
z-direction. The accelerated observer's trajectory in Minkowski
coordinates is a hyperbola that is indicated in terms of Rindler
coordinates $(\tau,\xi)$ \cite{Davies,Carroll}, with the following
form
 \bea\label{E1}
 (z,t)=\pm \left(\frac{{{e^{a\xi }}}}{a}\cosh (a\tau), \frac{{{e^{a\xi }}}}{a}\sinh (a\tau
 )\right),
   \eea
where $\tau$ is the Alice's proper time, $a$ is an arbitrary
reference acceleration and $\frac{{{e^{a\xi }}}}{a}$ is the proper
acceleration for Alice. The straight lines passing from origin are
obtained by the coordinate constant $\tau$, and hyperbola is
obtained by the coordinate $\xi$ as is plotted in Fig. \ref{Fig0}.
 The horizons $H_{\pm}$ that are
obtained by the light-like asymptotes, $z^{2}=t^{2}$, represent
proper times $\tau=\pm\infty$ in the limit of
$\xi\rightarrow-\infty$. The right half and the left half of
Minkowski plane are two regions that are called Rindler wedges I and
II, respectively. Alice and the fictitious observer, anti-Alice, are
constrained to move in the Rindler wedges I and II, respectively, as
these regions are causally disconnected from each other, i.e. no
information can propagate between them.

In a general discussion, we shall study superdense coding with an
accelerated particle in single mode approximation and beyond single
mode approximation. We cover the discussion in a general manner and
find the probability of success for superdense coding with uniformly
accelerated particle. We appraise the whole process by means of
superdense coding capacity, with definition given below. For the
sake of completeness, we also discuss the results in terms of
existing entanglement and quantum correlation and the corresponding
changes under superdense coding with uniformly accelerated particle.

Superdense coding capacity is the maximum value of classical
information that can be conveyed for a primary given state being
shared between Alice and Bob. When the encoding operator used in the
protocol is a unitary operator and the channel is noiseless, then
superdense coding capacity is defined as follow
\cite{Hiroshima,Bowen,Metwally}
  \bea\label{capacity}
  C(A:B)=\log_{2}d+S(\rho^{B})-S(\rho^{AB}).
  \eea
Here, $\rho^{B}$ is Bob's reduced density matrix, $\rho^{AB}$ is the
initial shared state and $d$ is the dimension of Alice's system.
$S(\rho)$ is the von Neumann entropy,
$S(\rho)=-\sum_{i}\lambda_{i}\log_{2}(\lambda_{i})$, where
$\lambda_{i}$'s are the eigenvalues of $\rho$.

Logarithmic negativity \cite{Peress,Horodeki} that is employed for
evaluating entanglement of $\rho$ is defined as
  \bea\label{N}
  N(\rho)=\log_{2}\sum_{i}|\lambda_{i}(\rho^{\rm pt})|,
  \eea
where $\lambda_{i}(\rho^{\rm pt})$'s are the eigenvalues of the
partial transpose of $\rho$.

Quantum discord is evaluated
\cite{ollivier,Vedral,Vedral2,saitoh,rahimi} for measuring
nonclassical correlation and it is defined as
 \bea\label{Discord}
   D(A:B)=\mathcal{I}(A:B)-\mathcal{C}(A:B),
 \eea
  where $\mathcal{I}(A:B)$ is quantum mutual information. It is determined as
 \bea\label{mutual}
   \mathcal{I}(A:B)=S(\rho^{A})+S(\rho^{B})-S(\rho^{AB}).
 \eea
 $\mathcal{C}(A:B)$ is the classical correlation given as follow
 \bea
   \mathcal{C}(A:B)=\mathop {\rm max }\limits_{\{ {\mathcal{B} _k}\} } [{\mathcal{J}_{\{ {\mathcal{B} _k}\} }}(A:B)] ,
 \eea
 where, $\mathcal{J}$ is locally accessible mutual information defined as follow
 \bea
 {\mathcal{J}_{\{ {\mathcal{B} _k}\} }}(A:B) = S({\rho _A}) - {S_{\{ {\mathcal{B}_k}\} }}(A|B).
 \eea
 ${S_{\{ {\mathcal{B} _k}\} }}(A|B)$ is the quantum conditional entropy defined as follow
 \bea\label{conditional}
 {S_{\{ {\mathcal{B} _k}\} }}(A|B)=\sum\limits_k {{p_k}S({\rho _{A|k}})},
 \eea
where $\{\rho_{k}, p_{k}\}$ is the ensemble of the outcome, after
von Neumann measurements $\{\mathcal{B} _k\}$ for the subsystem $B$,
and ${\rho _{A|k}} = {{\rm {Tr}}_B}({\mathcal{B}_k}\rho {\mathcal{B}
_k})/{p_k}$, with ${p_k} = {\rm Tr} ({\mathcal{B} _k}\rho
{\mathcal{B} _k})$. Calculating quantum discord for a general state
can be hard, however for special cases, e.g. where the state is a
X-state, there is a standard approach (see appendix A). The
resultant states being studied in the process of superdense coding
with uniformly accelerated particle are X-states. Therefore, we give
precise quantum discord values in addition to logarithmic negativity
values and compare them with superdense coding capacities.
\\
%*************************************************%%%%%%%%%%%%%%%%%%%%%%%%%%%%%%%%%%%
\section{Superdense coding in single-mode approximation}
%*************************************************%%%%%%%%%%%%%%%%%%%%%%%%%%%%%%%%%%%
  Considering a free Minkowski Dirac field in 1+1 dimensions, we assume all modes of
the field are in vacuum state except two modes that belong to Alice
and Bob. The Minkowski vacuum for Alice can be expanded in terms of
the corresponding Rindler vacuum \cite{Alsing}, as
 \bea
 &&\hspace{-.82 cm}|0\rangle_{A}= \cos r |0\rangle_{\rm I}|0\rangle_{\rm II} + \sin r |1\rangle_{\rm I}|1\rangle_{\rm II},\label{0}\\
 &&\hspace{-.82 cm}|1\rangle_{A}= |1\rangle_{\rm I}|0\rangle_{\rm II},\label{1}
 \eea
where $|i\rangle_{A}$ is the Minkowski particle mode belonging to
Alice, $|i\rangle_{\rm I}$ is the Rindler region I particle modes
and $|i\rangle_{\rm II}$ is the Rindler region II anti-particle
modes.

\begin{figure}
  \center
  \includegraphics[width=.5\columnwidth]{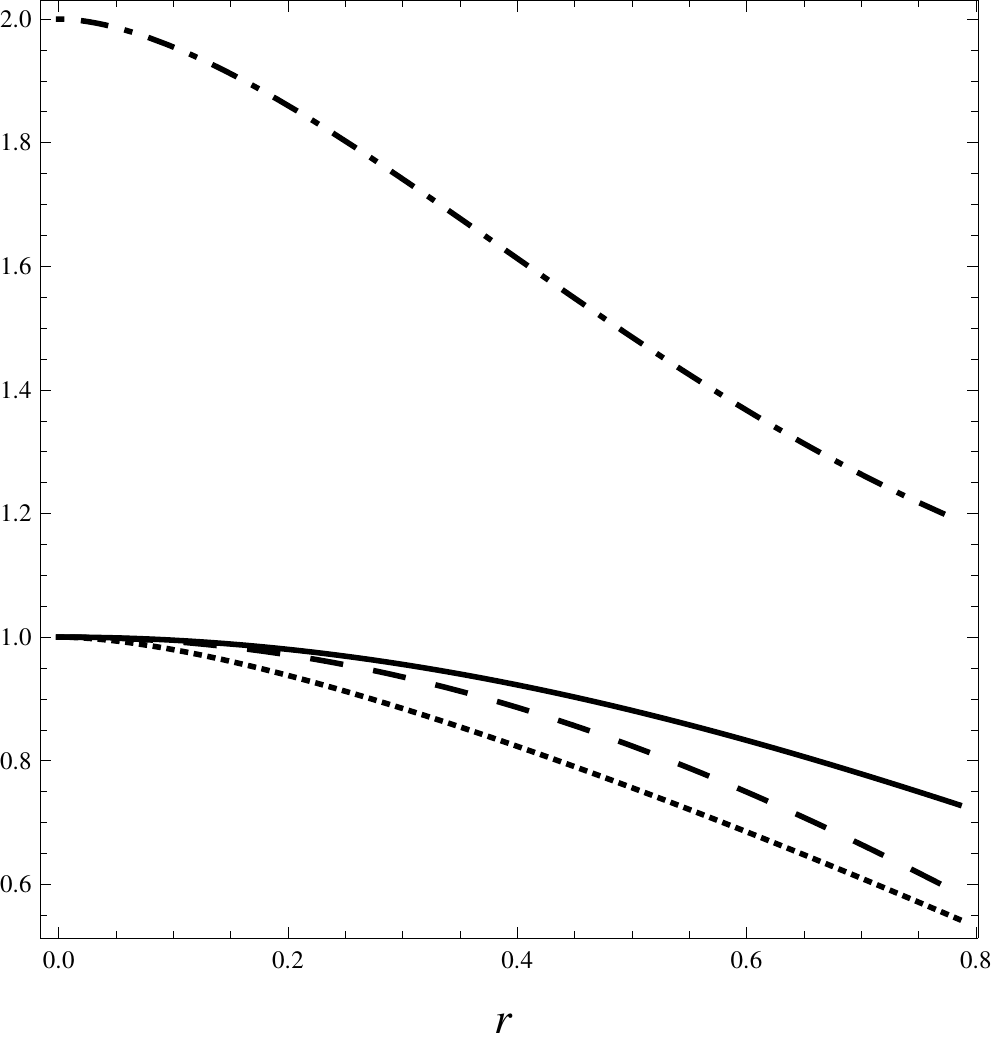}\\
  \caption{Probability of success for superdense coding,
   $P$, solid line, superdense coding capacity, $C({\rm I}:B)$, dotdashed line, logarithmic negativity, $N$, dashed line, and quantum discord, $D({\rm I}:B)$, dotted line,
   as functions of acceleration parameter, $r$, for $\rho_{00}^{{\rm I},B}$, in single mode approximation.}\label{Fig1}
\end{figure}

 In single mode approximation, the shared state $|\varphi_{00}\rangle_{AB}$ can be rewritten by substituting the relations Eq. (\ref{0}) and Eq. (\ref{1}) in Eq. (\ref{in}) only for Alice, as
\bea\label{00}
  |\varphi_{00}\rangle_{{\rm{I,II}},B}  = \frac{1}{\sqrt{2}} \left\{\cos r|000\rangle + \sin r |110\rangle +  |101\rangle\right\},
 \eea
where $|abc\rangle= |b\rangle_{\rm I} |c\rangle_{\rm
II}|a\rangle_{B} $. A unitary operator $U_{ij}$ is applied on Alice,
the accelerated qubit. The operator $I$ does not change the state
Eq. (\ref{00}), but other operators change the state into another
state, as follow
\begin{widetext}
 \bea\label{ij}
  %&&  \hspace{-1 cm}
  U_{ij}|\varphi_{00}\rangle_{{\rm{I,II}},B} & =& \frac{(-1)^{ij}}{\sqrt{2}} \left\{\cos r|i00\rangle + (-1)^{j} |\overline{i}01\rangle%\nonumber\\
  %&&  \hspace{3.2 cm}
  +(-1)^{j} \sin r |\overline{i}10\rangle
  \right\}\nonumber\\ %  &&\hspace{1.05 cm}
  &=&|\varphi_{ij}\rangle_{{\rm{I,II}},B}.
 \eea
  Then, the accelerated particle reaches Bob.
  If the information to be sent is $ij=00$ then the resultant density
matrix is as follow
  \bea
  %\hspace{-.84 cm}
  \rho_{00}^{{\rm{I,II}},B}&=&
  \frac{1}{2}
  \left\{\cos^{2}r|000\rangle\langle000|
  +\sin^{2}r|110\rangle\langle110|%\nonumber\\  &&\hspace{.4 cm}
  +|101\rangle\langle101|+\big(\cos r|000\rangle\langle101|\right.\nonumber\\ % \hspace{.4 cm}
  &&\hspace{.4 cm}\left. +\cos r\sin r|000\rangle\langle110|%\nonumber\\  &&\hspace{.4 cm}
  +
  \sin r|110\rangle\langle101|+{\rm h.c.}\big)\right\}.
  \eea
\end{widetext}
Recall that the Rindler regions I and II are causally
disconnected. Alice is constrained to move in region I, so by
tracing out region II, Bob obtains the shared density matrix, as
follow
  \bea\label{s00}
  \rho _{00}^{{\rm{I}},B} = \frac{1}{2}\left( {\begin{array}{*{20}{c}}
{{{\cos }^2}r}&0&0&{\cos r}\\
0&0&0&0\\
0&0&{{{\sin }^2}r}&0\\
{\cos r}&0&0&1
\end{array}} \right),
\eea where $|ab\rangle= |b\rangle_{\rm I}|a\rangle_{B} $. The
density matrix obtained for different cases of $ij$, the classical
message, can be found as
 \bea\label{st1}
 \rho_{ij}^{{\rm I},B}
  &=&{\rm Tr}_{\rm II}(\rho_{ij}^{{\rm I,II},B})%\nonumber
  \\
 & =&\frac{1}{2}\left\{\cos^{2}r|i0\rangle\langle i0|+\sin^{2}r|\bar{i}0\rangle\langle\bar{i}0|+|\bar{i}1\rangle\langle
 \bar{i}1|\right.
  \left.+(-1)^{j}\left(\cos r|i0\rangle\langle\bar{i}1|
    +\rm h.c.\right)\right\}.\nonumber
  \eea
 Eq. (\ref{st1}) represents four distinctive states that are X-states. For decoding the
classical message, a Bell basis measurement is performed to obtain
the following results
  \bea\label{p1}
  &&\langle\varphi_{ij}|\rho_{ij}^{{\rm I},B}|\varphi_{ij}\rangle=\frac{1}{4}(1+\cos r)^{2},\nonumber\\
  &&\langle\varphi_{i\bar{j}}|\rho_{ij}^{{\rm I},B}|\varphi_{i\bar{j}}\rangle=\frac{1}{4}(1-\cos r)^{2},\nonumber\\
  &&\langle\varphi_{\bar{i}j}|\rho_{ij}^{{\rm I},B}|\varphi_{\bar{i}j}\rangle= \langle\varphi_{\bar{i}\bar{j}}|\rho_{ij}^{{\rm I},B}|\varphi_{\bar{i}\bar{j}}\rangle%\nonumber\\
  =\frac{1}{4}\sin^{2} r.
 \eea
Results of this measurement on the density matrix, after tracing out
region {II}, is dependent on the acceleration parameter, $r$. In
other words, superdense coding is performed with a probability of
$r$. By letting $r=0$, corresponding to $a=0$, then superdense
coding is run absolutely in accordance with the original scenario
\cite{Bennentt}. Fig. \ref{Fig1} shows probability of success for
superdense coding, $P(\rho_{ij}^{{\rm
I},B})=\langle\varphi_{ij}|\rho_{ij}^{{\rm
I},B}|\varphi_{ij}\rangle$, Eq. (\ref{p1}), as a function of
acceleration parameter, $r$.
%\begin{figure}[t]
%     % Requires \usepackage{graphicx}
 %   \center
  %  \includegraphics[width=0.70\columnwidth]{fig1.eps}\\
  % \caption{????????????}%\label{Fig1}
  %\end{figure}

In order to  evaluate superdense coding capacity and later for
quantum discord, we need to calculate the von Neumann entropies as
follows
   \bea\label{entropy1}
  &&S(\rho^{{\rm I},B})= - \frac{{1 - \cos 2r}}{4}\log_{2}\frac{1-\cos2r}{4}
   - \frac{{3 + \cos 2r}}{4}\log_{2}\frac{3+\cos2r}{4},\nonumber\\
  &&\hspace{0.3 cm} S(\rho^{{\rm I}})=- \frac{\cos^2 r}{2}\log_{2}\frac{\cos^2 r}{2} - \frac{1+\sin^2 r}{2}\log_{2}\frac{1+\sin^2 r}{2},\nonumber\\
  &&\hspace{0.2 cm} S(\rho^{B})=1.
   \eea
  Thus, superdense coding capacity, Eq. (\ref{capacity}), for the state Eq. (\ref{s00}), is obtained as follow
  \bea\label{capacity1}
  C({\rm I}:B)=2+ \frac{{1- \cos 2r}}{4}\log_{2}\frac{1-\cos2r}{4}
   + \frac{{3 + \cos 2r}}{4}\log_{2}\frac{3 +\cos2r}{4}.
   \eea
   %Fig. \ref{Fig4}. shows the behavior of Eq.  (\ref{ca}) as function of $r$.
In Fig. \ref{Fig1}, superdense coding capacity, Eq.
(\ref{capacity1}), is plotted as a function of acceleration
parameter, $r$.

 Logarithmic negativity, Eq. (\ref{N}), is calculated for the entanglement of the state, Eq. (\ref{s00}). Eigenvalues of the partial
transpose of the density matrix $\rho_{00}^{{\rm I},B}$, are given
by
  \bea
  &&%\hspace{-1.2 cm}
  \lambda_{1,2}(\rho^{\rm pt}_{{\rm I},B})=\frac{1}{2},\nonumber\\
  &&%\hspace{-1 cm}
  \lambda_{3,4}(\rho^{\rm pt}_{{\rm I},B})= \pm \frac{\cos^{2}r}{2}.
  \eea
 Thus, logarithmic negativity can be written as follow
  \bea\label{N1}
  N(\rho _{00}^{{\rm I},B}) = {\log _2}\left(1 + \cos^{2}r\right).
   \eea
  Fig. \ref{Fig1} indicates Eq. (\ref{N1}) as a function of $r$.

Quantum discord is given by Eq. (\ref{Discord}).
   % \bea\label{discord}
%    D({\rm I}:B)=\mathop {\min }\limits_{\left\{ {{\mathcal{B}_k}} \right\}} S_{\{\mathcal{B}_{k}\}}({\rm I}|B)+S(\rho^{{\rm I}})-S(\rho^{{\rm I},B}).
%    \eea
 For the state of Eq. (\ref{s00}), after evaluating the corresponding von Neumann entropies, Eqs. (\ref{entropy1}), and employing the approach explained in Refs. \cite{Ali,Chen},
 %appendix A,
 quantum discord is calculated for which Fig. \ref{Fig1} shows its behavior as a function of $r$.
 It is clear, that four quantities, probability of success, superdense coding capacity, logarithmic negativity and quantum discord for superdense coding with accelerated particle, in single mode approximation, are descending functions of $r$.

%*************************************************%%%%%%%%%%%%%%%%%%%%%%%%%%%%%%%%%%%
\section{Superdense coding in beyond single-mode approximation}
\begin{figure}
  \includegraphics[width=0.5\columnwidth]{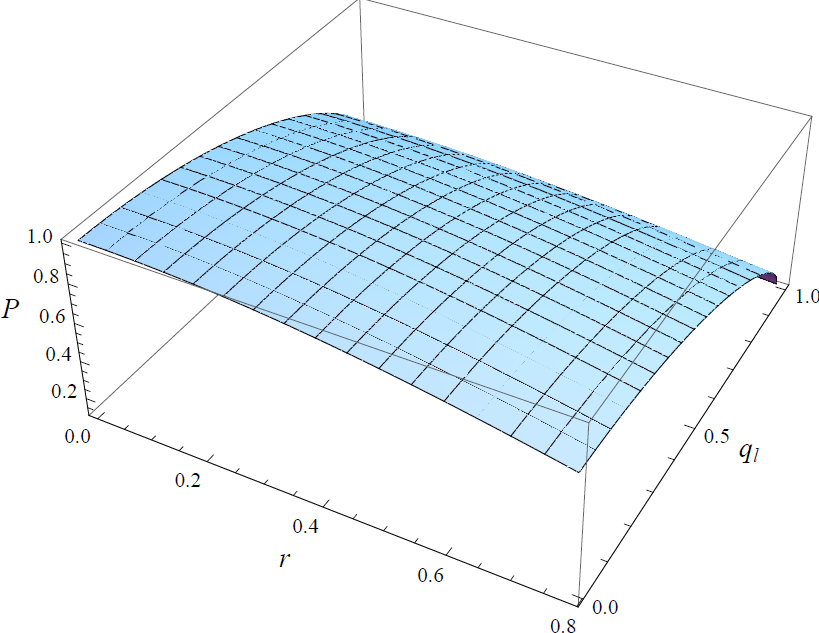}\\
  \caption{Probability of success in terms of acceleration parameter, $r$, and presence possibility of the particle in region II of Rindler region, $q_{l}$.}\label{Fig4}
\end{figure}

%*************************************************%%%%%%%%%%%%%%%%%%%%%%%%%%%%%%%%%%%
 In beyond single mode approximation, an accelerated
detector can detect a mode in both Rindler wedges I and II,
therefore there are different right and left components for the
single-particle state denoted as Alice \cite{Bruschi},
 \bea\label{by}
 &&%\hspace{-.82 cm}
 |0\rangle_{A}= \cos r |0\rangle_{\rm I}|0\rangle_{\rm II} + \sin r |1\rangle_{\rm I}|1\rangle_{\rm II},\nonumber\\
 &&%\hspace{-.82 cm}
 | 1\rangle_{A}= {q_l}{\left| 0 \right\rangle _{\rm
I}}{\left| 1 \right\rangle _{\rm II}} + {q_r}{\left| 1 \right\rangle
_{\rm I}}{\left| 0 \right\rangle _{\rm II}},
 \eea
where ${q_l}$ and ${q_r}$ are complex numbers that satisfy
$q_{l}^{2}+q_{r}^{2}=1$. For simplicity, we only consider the cases
that ${q_l}$ and ${q_r}$ are real. The single mode approximation is
found by letting $ q_r = 1$ in the general form Eq. (\ref{by}). The
shared state $ |\varphi_{00}\rangle_{AB}$ can be rewritten by
substituting the relations Eq. (\ref{0}) and Eq. (\ref{by}) for
Alice in Eq. (\ref{in}), in beyond single mode approximation, as
 \bea
   |\varphi_{00}\rangle_{{\rm{I,II}},B}
    &=& \frac{1}{\sqrt{2}} \left\{\cos r |000\rangle + \sin r |110\rangle\right.
    \left.+ q_{l} |011\rangle + q_{r} |101\rangle\right\}.\label{q00}
 \eea
  Alice applies a unitary operator $U_{ij}$ on her qubit. Like the previous section,
operator $I$ does not change the state Eq. (\ref{q00}), but others
do change the state into another state, as follows
\begin{widetext}
\bea\label{qij}
   U_{ij}|\varphi_{00}\rangle_{{\rm{I,II}},B}
  & =& \frac{(-1)^{ij}}{\sqrt{2}} \left\{\cos r |i00\rangle+ (-1)^{j} \sin r |\overline{i}10\rangle%\right.  \nonumber\\  &&\hspace{1.2 cm}\left.
   +q_{l} |i11\rangle
    + (-1)^{j}q_{r}|\overline{i}01\rangle\right\}
    \nonumber\\
  & =&|\varphi_{ij}\rangle_{{\rm{I,II}},B}.
  \eea
 Now, the state in Bob's possession, after he receives the accelerated particle is  $\left| \varphi_{ij}
\right\rangle_{{\rm{I,II}},B}$. For the case $ij=00$, the resultant
density matrix is given by
  \bea
  \rho_{00}^{{\rm{I,II}},B}&&\hspace{-0.4 cm}=%&=&
  \frac{1}{2}
  \Big\{\cos^{2}r|000\rangle\langle000|+\sin^{2}r|110\rangle\langle110|%\nonumber\\  &&\hspace{.9 cm}
  +q_{l}^{2}|011\rangle\langle011|+q_{r}^{2}|101\rangle\langle101|%\nonumber\\  &&\hspace{.4 cm}
  +\Big(\cos r\sin r|000\rangle\langle110|\\  &&\hspace{0.4 cm}
  +q_{l}\sin r |110\rangle\langle011|+q_{r}\cos r|000\rangle\langle101|%\nonumber\\  &&\hspace{.4 cm}
  +q_{l}\cos r |000\rangle\langle011|
    +q_{r}\sin r |110\rangle\langle101|+q_{l} q_{r} |011\rangle\langle101|%\nonumber\\  &&\hspace{1.1 cm}
  +\rm h.c.\Big)\Big\}.\nonumber%\\
  \eea
  The density matrix that is given by tracing out region
II, is given by
  \bea\label{st00}
 %\hspace{-1 cm}
 \rho_{00}^{{\rm I},B}%&=&
 ={\rm Tr}_{{\rm II}}(\rho_{00}^{{\rm I,II},B})%\nonumber\\ \hspace{-.3 cm}&=&
 =\frac{1}{2}\left( {\begin{array}{*{20}{c}}
{{{\cos }^2}r}&0&0&{{q_r}\cos r}\\
0&{q_l^2}&{{q_l}\sin r}&0\\
0&{{q_l}\sin r}&{{{\sin }^2}r}&0\\
{{q_r}\cos r}&0&0&{q_r^2}
\end{array}} \right).
  \eea
  For other cases for the classical message $ij$, the density matrix can be obtained as follow
  \bea\label{st}
  \rho_{ij}^{{\rm I},B}
 & =&{\rm Tr}_{\rm II}(\rho_{ij}^{{\rm I,II},B})\\
 & =&\frac{1}{2}\Big\{\cos^{2}r|i0\rangle\langle i0|+\sin^{2}r|\bar{i}0\rangle\langle\bar{i}0|%\nonumber\\ &&\hspace{.1 cm}
 +q_{l}^{2}|i1\rangle\langle i1|+q_{r}^{2}|\bar{i}1\rangle\langle \bar{i}1|%\nonumber\\&&
 %\hspace{.1 cm}
  +(-1)^{j}\Big(q_{r}\cos r|i0\rangle\langle\bar{i}1|+ q_{l}\sin r|\bar{i}0\rangle\langle i1|%\nonumber\\ &&\hspace{1.45 cm}
 +\rm h.c.\Big)\Big\},\nonumber
  \eea
  \end{widetext}
which represents four distinctive matrices that are X-forms. Thus,
measurement in Bell basis by Bob yields
   \bea\label{p2}
  &&\hspace{-1 cm}\langle\varphi_{ij}|\rho_{ij}^{{\rm I},B}|\varphi_{ij}\rangle=\frac{1}{4}(q_{r}+\cos r)^{2},\nonumber\\
  &&\hspace{-1 cm}\langle\varphi_{i\bar{j}}|\rho_{ij}^{{\rm I},B}|\varphi_{i\bar{j}}\rangle=\frac{1}{4}(q_{r}-\cos r)^{2},\nonumber \\
  &&\hspace{-1 cm}\langle\varphi_{\bar{i}j}|\rho_{ij}^{{\rm I},B}|\varphi_{\bar{i}j}\rangle=\frac{1}{4}(q_{l}+\sin r)^{2},\nonumber\\
  &&\hspace{-1 cm}\langle\varphi_{\bar{i}\bar{j}}|\rho_{ij}^{{\rm I},B}|\varphi_{\bar{i}\bar{j}}\rangle=\frac{1}{4}(q_{l}-\sin r)^{2}.
  \eea
These results show the probability of success, $P(\rho_{ij}^{{\rm
I},B})$, is $\frac{1}{4}(q_{r}+\cos r)^{2}$, and it is illustrated
in Fig. \ref{Fig4}, Fig.  \ref{Fig2} and Fig. \ref{Fig3}. Therefore, measurement by
Bob depends on the acceleration parameter, $r$. If $r=0$ and
$q_{r}=1$, corresponding to $a=0$ and $q_{l}=0$, respectively, then
the original superdense coding scenario is given \cite{Bennentt}.

\begin{figure}
 \includegraphics[width=0.5\columnwidth]{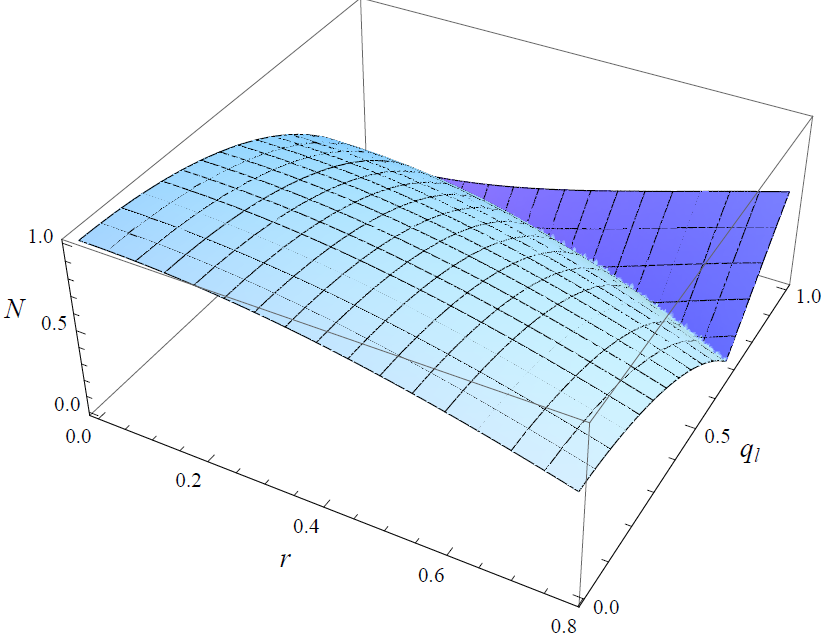}\\
  \caption{Capacity of superdense coding in terms of acceleration parameter, $r$, and presence possibility of the particle in region II of Rindler region, $q_{l}$.}\label{Fig5}
\end{figure}

For the state Eq. (\ref{st00}), the von Neumann entropies are given
as follows
   \bea\label{entropy}
   %&&
   S(\rho^{{\rm I},B})=&&\hspace{-.3cm} - \frac{{-1 + 2q_l^2 - \cos 2r}}{4}\log_{2}\frac{-1 + 2q_{l}^{2}-\cos2r}{4}- \frac{{3 - 2q_l^2 + \cos 2r}}{4}\log_{2}\frac{3 - 2q_{l}^{2}+\cos2r}{4},\nonumber\\
     %&&
     S(\rho^{{\rm I}})=&&\hspace{-.3cm}- \frac{{1-q_l^2 + {{\sin^2 r}}}}{2}\log_{2}\frac{1-q_{l}^{2}+\sin^2 r}{2}- \frac{{q_l^2 + {{\cos^2 r}}}}{2}\log_{2}\frac{q_{l}^{2}+\cos^2
    r}{2},\nonumber\\
     %&&
     S(\rho^{B})=&&\hspace{-.3cm}1.
   \eea
   Thus, superdense coding capacity $C({\rm I}:B)$, Eq. (\ref{capacity}), is calculated as follow
  \bea\label{ca}
 % &&\hspace{-1 cm}
 C({\rm I}:B)=%&=&
 &&\hspace{-.35cm}2+ \frac{{3 - 2q_l^2 + \cos 2r}}{4}\log_{2}\frac{3-2q_{l}^{2}+\cos2r}{4} + \frac{{1 + 2q_l^2 - \cos
2r}}{4}\log_{2}\frac{1+2q_{l}^{2}-\cos2r}{4}.
   \eea
  Fig. \ref{Fig5}, Fig. \ref{Fig2} and Fig. \ref{Fig3} show the behavior of Eq.  (\ref{ca}) as a function of $r$ and $q_{l}$, respectively.

\begin{figure}
  \includegraphics[width=0.5\columnwidth]{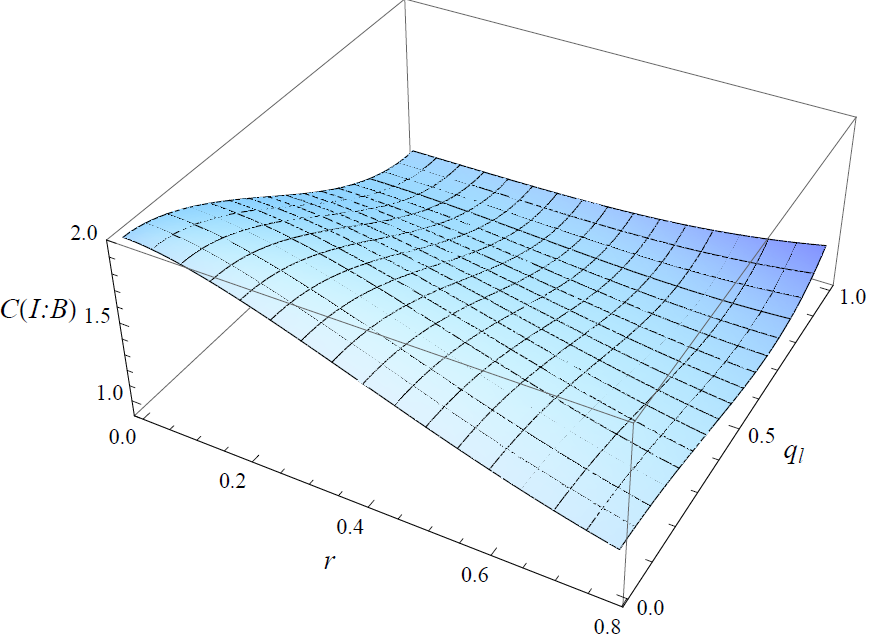}\\
  \caption{Logarithmic negativity in terms of acceleration parameter, $r$, and presence possibility of the particle in region II of Rindler region, $q_{l}$.}\label{Fig6}
\end{figure}

    The entanglement of Eq. (\ref{st00}) is evaluated by logarithmic negativity, Eq. (\ref{N}). Eigenvalues of the partial transpose of the density matrix $\rho_{00}^{{\rm I},B}$, are given by
  \bea
  &&\lambda_{1,2}(\rho^{\rm pt}_{{\rm I},B})=\frac{1}{2},\nonumber\\
%  &&\lambda_{3}(\rho^{\rm pt}_{{\rm I},B})= + \frac{1}{2}\left( {\cos^{2}r - {q_l}^2} \right),\nonumber\\
  &&\lambda_{3,4}(\rho^{\rm pt}_{{\rm I},B})= \pm \frac{1}{2}\left( {\cos^{2}r - {q_l}^2} \right).
  \eea
 Thus, logarithmic negativity is calculated as follow
  \bea
  N(\rho _{00}^{{\rm I},B}) = {\log _2}\left(1+ \left| \cos^{2}r - q_l^2 \right|\right).
   \eea
 Fig. \ref{Fig6}, Fig. \ref{Fig2} and Fig. \ref{Fig3} show the behavior of Eq. (\ref{capacity}) as a function of $r$ and $q_{l}$, respectively.

\begin{figure}
  \includegraphics[width=0.5\columnwidth]{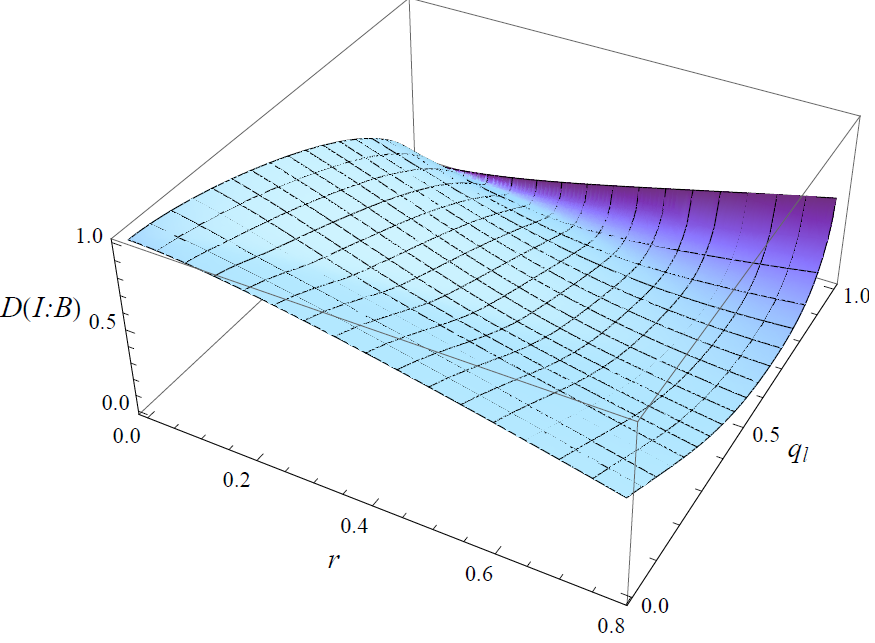}\\
  \caption{Quantum discord in terms of acceleration parameter, $r$, and presence possibility of the particle in region II of Rindler region, $q_{l}$.}\label{Fig7}
\end{figure}

 Quantum discord, Eq. (\ref{Discord}), is derived by considering the corresponding von Neumann entropies,  Eqs. (\ref{entropy}), and following the approach in Refs. \cite{Ali,Chen}.
 % appendix A.
 Fig. \ref{Fig7}, Fig. \ref{Fig2} and Fig. \ref{Fig3} show nonclassical correlation in terms of $r$ and $q_{l}$, respectively.

%%%%%%%%%%%%%%%%%%%%%%%%%%%%%%%%%%%%%%%%%%%%%%%%
\section{Generality of discussions for all $\alpha,\beta=\{0,1\}$, Eq. (\ref{in})}%\appendix{\textbf{Appendix B : General form}}
%%%%%%%%%%%%%%%%%%%%%%%%%%%%%%%%%%%%%%%%%%%%%%%
Generally, the initial shared state $ |\varphi_{\alpha\beta}\rangle_{A,B}$ can be rewritten by substituting
the relations Eq. (\ref{0}) and Eq. (\ref{by}) in beyond single mode
approximation, as
 \bea\label{qkl}
   |\varphi_{\alpha\beta}\rangle_{{\rm{I,II}},B}
    = \frac{1}{\sqrt{2}} \left\{\cos r |00\alpha\rangle + \sin r |11\alpha\rangle %\nonumber\\
    + (-1)^{\beta}q_{l} |01\bar{\alpha}\rangle +(-1)^{\beta} q_{r} |10\bar{\alpha}\rangle\right\}.\nonumber\\
 \eea
A unitary operator $U_{ij}$ is applied on the accelerated particle.
Then the resultant state is sent to Bob. The operator $I$ does
  not change the state Eq. (\ref{qkl}), but others do change the state
into another state, as follow \bea\label{qijkl}
  %&&\hspace{-1 cm}
   U_{ij}|\varphi_{\alpha\beta}\rangle_{{\rm{I,II}},B}
   &=& \frac{(-1)^{ij}}{\sqrt{2}} \left\{\cos r |i0\alpha\rangle+(-1)^{\beta} q_{l} |i1\bar{\alpha}\rangle % \nonumber\\  &&\hspace{1.4 cm}
   +(-1)^{j} \sin r |\overline{i}1\alpha\rangle
    %\right.\nonumber\\  &&\hspace{1.2 cm}\left.
    + (-1)^{\beta+j}q_{r}|\overline{i}0\bar{\alpha}\rangle\right\}.
   \eea
 This is the state in Bob's possession. The resultant density matrix for superdense coding beyond single mode approximation is given by tracing out region
II, as follow
  \bea\label{stijkl}
 &&\hspace{-1 cm} \rho^{{\rm I},B}
  ={\rm Tr}_{\rm II}(\rho^{{\rm I,II},B})\\
 &&\hspace{-.3 cm} =\frac{1}{2}\Big\{\cos^{2}r|i\alpha\rangle\langle i \alpha|+\sin^{2}r|\bar{i}\alpha\rangle\langle\bar{i}\alpha|%\nonumber\\ &&\hspace{.6 cm}
 +q_{l}^{2}|i\bar{\alpha}\rangle\langle i\bar{\alpha}|+q_{r}^{2}|\bar{i}\bar{\alpha}\rangle\langle \bar{i}\bar{\alpha}|\nonumber\\ &&\hspace{.6 cm}
  +(-1)^{\beta+j}\left(q_{r}\cos r|i\alpha\rangle\langle\bar{i}\bar{\alpha}|+ q_{l}\sin r|\bar{i}\alpha\rangle\langle i\bar{\alpha}|%\nonumber\\ &&\hspace{2.35 cm}
  +\rm h.c.\right)\Big\},\nonumber
  \eea
which represents X-form matrices for all cases of
$\alpha,\beta,i,j$.

%%%%%%%%%%%%%%%%%%%%%%%%%%%%%%%%%%%%%%%%%%%%%%%%%%%%%%%%%%%%%%%%%%%%%%%%%%%%%%%%

 %It is noteworthy that
 Therefore, our initial assumption of $\alpha=\beta=0$ for the shared entanglement, Eq. (\ref{in}), does not affect the generality of discussions for single mode approximation and beyond single mode approximation. For both of the cases, the resultant states from superdense coding with uniformly accelerated particle can be evaluated for their probabilities of success, superdense coding capacities, negativity values and discord values. The final states, for all four choices of $\alpha$ and $\beta$, are X-form states.
 %see appendix B.
 Therefore, quantum discord can be calculated following the approach discussed in Refs. \cite{Ali,Chen}.
%%%%%%%%%%%%%%%%%%%%%%%%%%%%%%%%%%%%%%%%%%%%%%%%
\section{Discussions and Conclusion}
%%%%%%%%%%%%%%%%%%%%%%%%%%%%%%%%%%%%%%%%%%%%%%%%
%\begin{figure}[t]
%     % Requires \usepackage{graphicx}
%    \center
%    \includegraphics[width=0.70\columnwidth]{fig1.eps}\\
%    \caption{????????????}\label{Fig2}
%   \end{figure}
\begin{figure}
  \center
  \includegraphics[width=.5\columnwidth]{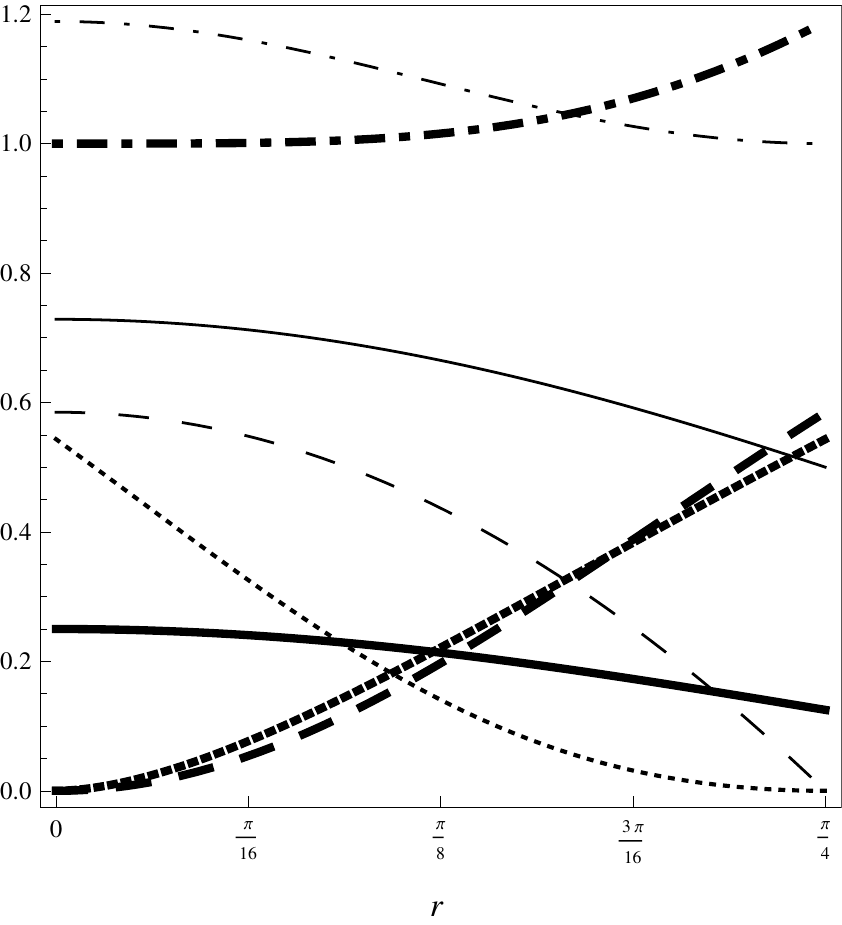}\\
  \caption{Probability of success for superdense coding,
$P$, solid lines, superdense coding capacity, $C({\rm I}:B)$,
dotdashed lines, logarithmic negativity, $N$, dashed lines, and
quantum discord, $D({\rm I}:B)$, dotted lines, for
$q_{l}=\frac{1}{\sqrt{2}}$, thin lines, and $q_{l}=1$, thick lines,
as functions of $r$, for $\rho_{00}^{{\rm I},B}$, in beyond single
mode approximation.}\label{Fig2}
\end{figure}
\begin{figure}
  \center
  \includegraphics[width=.5\columnwidth]{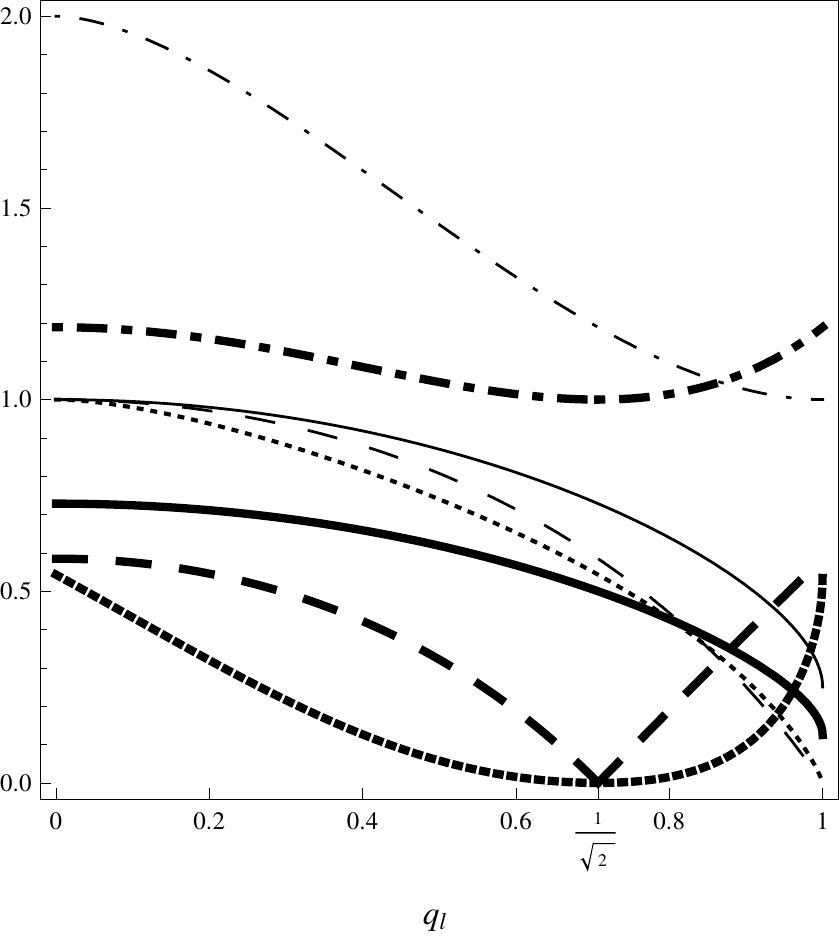}\\
  \caption{Probability of success for superdense coding,
$P$, solid lines, superdense coding capacity, $C({\rm I}:B)$,
dotdashed lines, logarithmic negativity, $N$, dashed lines, and
quantum discord, $D({\rm I}:B)$, dotted lines, for $r=0$, thin
lines, and $r=\frac{\pi}{4}$, thick lines, as functions of $q_{l}$,
for $\rho_{00}^{{\rm I},B}$, in beyond single mode
approximation.}\label{Fig3}
\end{figure}
We studied superdense coding with uniformly accelerated particle in
single mode approximation and beyond single mode approximation. In
single mode approximation, $q_{r}=1$ (or equally $q_{l}=0$),
measurement by Bob on the density matrix after tracing out region
{II} is dependent on the acceleration parameter, $r$. By letting
$r=0$, corresponding to $a=0$, superdense coding is performed with
absolute probability, Eq. (\ref{p1}), in accordance with the
original superdense coding \cite{Bennentt}. As illustrated in Fig.
\ref{Fig1}, probability of success, superdense coding capacity,
logarithmic negativity and quantum discord are all descending
functions of acceleration parameter, $r$.

In beyond single mode approximation, the situation is more
intricate. Fig. \ref{Fig2} (Fig. \ref{Fig3}) is to show behaviors of
probability of success, superdense coding capacity, negativity and
quantum discord for the resultant state of superdense coding with
uniformly accelerated particle, for distinct values of $q_{l} ~(r)$,
as functions of $r ~(q_{l})$. $q_{l}$ is in interval [0,1]. Fig.
\ref{Fig2} shows the functions for $q_{l}$ maximum that is
$q_{l}=1$, and for $q_{l}=\frac{1}{\sqrt{2}}$. Entanglement and
nonclassical correlation are zero for $q_{l}=\frac{1}{\sqrt{2}}$,
with $r=\frac{\pi}{4}$. In Fig. \ref{Fig3}, the four functions are
shown for $r$ minimum, that is $r=0$, and for $r=\frac{\pi}{4}$,
that is when quantum correlations are zero at
$q_{l}=\frac{1}{\sqrt{2}}$.

Recall that single mode approximation is a special case for beyond
single mode approximation for when $q_{l}=0$. From Fig. \ref{Fig3}
,we can see that for $q_{l}=0$, and two cases of $r=0$ and
$r=\frac{\pi}{4}$, the evaluated functions values exactly coincide
with the corresponding ones being represented in Fig. \ref{Fig1}.

In Fig. \ref{Fig2}, when $q_{l}=1$, the maximum value for $q_{l}$,
the maximum probability of success, $P$, is for $r=0$, Eq.
(\ref{p2}). $P$ is decreasing with increasing $r$. We would expect
similar behaviors for entanglement, nonclassical correlation and the
capacity, however negativity and discord, as well as the capacity,
are representing increasing behaviors. In beyond single mode
approximation, Eq. (\ref{by}), if the accelerated object starts from
$|1\rangle$, there is some distinct probability for the state to
change to $|0\rangle$, and this probability is equal to $1$
specifically for when $q_{l}=1$, the case illustrated in Fig.
\ref{Fig2} with thick lines. Indeed, we do not evaluate the
entanglement, nor nonclassical correlation of the original shared
entangled state by negativity and discord, and what is illustrated
is actually the negativity and discord for the state
$|\psi_{\bar{i}j}\rangle$, but not the original state
$|\psi_{ij}\rangle$. The same discussion is applied to explain the
capacity of superdense coding since this function is evaluated using
nonclassical correlations. We, therefore, conclude that the
probability of success is the best means for evaluating the process
of superdense coding with accelerated particle, specially for a
large $q_{l}$, i.e. when beyond single mode approximation is
strongly used.

In Fig. \ref{Fig2}, when $q_{l}=\frac{1}{\sqrt{2}}$, since $q_{l}$
is not very large, i.e. even in beyond single mode approximation,
the initial state of the accelerated particle only changes to an
unbiased superposition of $|0\rangle$ and $|1\rangle$, Eq.
(\ref{by}). Therefore, we do not see any peculiar behavior from the
studied functions, as the previous paragraph. Here, the capacity of
superdense coding, entanglement, discord and the probability of
success are all decreasing functions with regard to $r$.

In Fig. \ref{Fig3}, when $r=0$, with an increase in $q_{l}$, the
four evaluated functions decrease, which is the expected behavior,
consulting the corresponding equations, and specifically Eq.
(\ref{by}). In the same figure, when $r=\frac{\pi}{4}$, with an
increase in $q_{l}$, entanglement and nonclassical correlation
decrease until they reach the minimum value $\frac{1}{\sqrt{2}}$.
From this point, the behaviors of these two functions are changed.
They represent increasing behaviors, which can be explained again by
Eq. (\ref{by}), since the state $|\psi_{ij}\rangle$ changes to
$|\psi_{\bar{i}j}\rangle$. Correspondingly, the capacity of
superdense coding is showing similar peculiar behavior. The capacity
of superdense coding generally follows the behavior of quantum
correlations, however the relationship is not as simple to give an
exact form. The probability of success is presenting behavior as the
expectation.

In relativistic regimes, superdense coding with an accelerated
particle and its probability of success can be reliably used for
evaluating the involved quantum states in terms of their
capabilities for being employed and manipulated for quantum
information processing purposes. In this regard, negativity, discord
and superdense coding capacity definitions are shown to have
obstacles at least for specific ranges of acceleration and in a
general form where one investigates the process in a general manner,
i.e. in beyond single mode approximation.

\end{document}